\documentclass[manuscript,nonacm]{acmart}

\AtBeginDocument{%
  \providecommand\BibTeX{{%
    \normalfont B\kern-0.5em{\scshape i\kern-0.25em b}\kern-0.8em\TeX}}}

\setcopyright{acmcopyright}
\copyrightyear{2018}
\acmYear{2018}
\acmDOI{XXXXXXX.XXXXXXX}

\usepackage{xparse}
\NewDocumentCommand{\etal}{ssO{et al. }}{%
  \IfBooleanTF{#1}{%
    {%
     \bfseries\itshape #3}%
    }{%
      \textit{#3}%
    }%
}
\NewDocumentCommand{\eg}{ssO{e.g., }}{%
  \IfBooleanTF{#1}{%
    {%
     \bfseries\itshape #3}%
    }{%
      \textit{#3}%
    }%
}
\NewDocumentCommand{\ie}{ssO{i.e., }}{%
  \IfBooleanTF{#1}{%
    {%
     \bfseries\itshape #3}%
    }{%
      \textit{#3}%
    }%
}
     
\providecommand{\DIFdel}[1]{} 

\usepackage[inline]{enumitem}
\usepackage{graphicx}
\usepackage{subcaption}
\usepackage{acro}
\usepackage[flushleft]{threeparttable}
\usepackage{pdflscape}
\usepackage{tabularx}
\usepackage{multirow}
\usepackage{afterpage}
\usepackage{adjustbox}
\usepackage[graphicx]{realboxes}
\usepackage{wrapfig}
\usepackage{lipsum}
\usepackage{rotating}
\usepackage{needspace}
\usepackage{longtable}
\usepackage{tabularx}
\usepackage{ltablex}
\usepackage{xcolor}

\DeclareAcronym{hais}{
  short = HAIs,
  long  = Healthcare-Associated Infections
}

\DeclareAcronym{who}{
  short = WHO,
  long  = World Health Organization
}

\DeclareAcronym{hcws}{
  short = HCWs,
  long  = Healthcare Workers
}

\DeclareAcronym{xgboost}{
  short = XGBoost,
  long  = eXtreme Gradient Boosting
}

\DeclareAcronym{semg}{
  short = sEMG,
  long  = Surface Electromyography
}

\DeclareAcronym{imu}{
  short = IMU,
  long  = Inertial Measurement Unit}

\DeclareAcronym{rf}{
  short = RF,
  long  = Random Forest
  }

\DeclareAcronym{svm}{
  short = SVM,
  long  = Support Vector Machine}

\DeclareAcronym{hmm}{
  short = HMM,
  long  = Hidden Markov Model
  }

\DeclareAcronym{loso}{
  short = LOSO,
  long  = Leave-One-Session-Out}

\DeclareAcronym{lopo}{
  short = LOPO,
  long  = Leave-One-Participant-Out}

\DeclareAcronym{rfid}{
  short = RFID,
  long  = Radio-frequency Identification}

\DeclareAcronym{uv}{
  short = UV,
  long  = Ultraviolet}

\DeclareAcronym{mp4}{
  short = MP4,
  long = MPEG-4 Part 14}

\DeclareAcronym{cnc}{
  short = CNC,
  long = Computer numerical control}

\DeclareAcronym{rgb}{
  short = RGB,
  long = Red Green Blue}

\DeclareAcronym{rtls}{
  short = RTLS,
  long = Real-time Locating System}

\DeclareAcronym{ble}{
 short = BLE,
  long = Bluetooth Low Energy}
 
\DeclareAcronym{icu}{
  short = ICU,
  long = intensive care unit}

\DeclareAcronym{rfi}{
  short = RFI,
  long = Radio-frequency Interference}

\DeclareAcronym{emi}{
  short = EMI,
  long = Electromagnetic Interference}

\DeclareAcronym{iot}{
  short = IoT,
  long = Internet of Things}

\DeclareAcronym{led}{
  short = LED,
  long = Light-Emitting Diode}

\DeclareAcronym{gui}{
  short = GUI,
  long = Graphical User Interface}

\DeclareAcronym{ar}{
  short = AR,
  long = Augmented Reality}

\DeclareAcronym{vr}{
  short = VR,
  long = Virtual Reality}

\DeclareAcronym{mr}{
  short = MR,
  long = Mixed Reality}

\DeclareAcronym{pc}{
  short = PC,
  long = Personal Computer}

\DeclareAcronym{prisma}{
  short = PRISMA,
  long = Preferred Reporting Items for Systematic Reviews and Meta-Analysis}

\DeclareAcronym{cnn}{
  short = CNN,
  long = Convolutional Neural Network}

\DeclareAcronym{hsv}{
  short = HSV,
  long = Hue\, Saturation\, Value}

\DeclareAcronym{roi}{
  short = ROI,
  long = Region of Interest}

\DeclareAcronym{csv}{
  short = CSV,
  long = Comma-Separated Values}

\DeclareAcronym{tp}{
  short = TP,
  long = True\,Positive}

\DeclareAcronym{fp}{
  short = FP,
  long = False\,Positive}

\DeclareAcronym{fn}{
  short = FN,
  long = False\,Negative}

\DeclareAcronym{tn}{
  short = TN,
  long = True\,Negative}

\DeclareAcronym{anova}{
  short = ANOVA,
  long = Analysis of Variance}

\DeclareAcronym{ols}{
  short = OLS,
  long = Ordinary Least Squares}

\begin{document}

\title{Using Thermal Imaging to Measure Hand Hygiene Quality}

\author{Chaofan Wang}
\email{chaofanw@student.unimelb.edu.au}
\affiliation{%
  \institution{Delft University of Technology}
  \city{Delft}
  \state{South Holland}
  \country{Netherlands}
  \postcode{2628 CD}
}

\author{Weiwei Jiang}
\affiliation{%
  \institution{Anhui Normal University}
  \city{Carlton}
  \state{Victoria}
  \country{China}
  \postcode{3000}
}

\author{Kangning Yang}
\affiliation{%
  \institution{The University of Melbourne}
  \city{Carlton}
  \state{Victoria}
  \country{Australia}
  \postcode{3000}
}

\author{Zhanna Sarsenbayeva}
\affiliation{%
  \institution{The University of Sydney}
  \city{Camperdown}
  \state{New South Wales}
  \country{Australia}
  \postcode{2006}
}

\author{Benjamin Tag}
\affiliation{%
  \institution{Monash University}
  \city{Carlton}
  \state{Victoria}
  \country{Australia}
  \postcode{3000}
}

\author{Tilman Dingler}
\affiliation{%
  \institution{The University of Melbourne}
  \city{Carlton}
  \state{Victoria}
  \country{Australia}
  \postcode{3000}
}

\author{Jorge Goncalves}
\affiliation{%
  \institution{The University of Melbourne}
  \city{Carlton}
  \state{Victoria}
  \country{Australia}
  \postcode{3000}
}

\author{Vassilis Kostakos}
\affiliation{%
  \institution{The University of Melbourne}
  \city{Carlton}
  \state{Victoria}
  \country{Australia}
  \postcode{3000}
}

\renewcommand{\shortauthors}{Wang \etal}

\begin{abstract}
\begin{description}[leftmargin=0cm]
    \item[Objectives:] Hand hygiene has long been promoted as the most effective way to prevent the transmission of infection. However, due to the low compliance and quality of hand hygiene reported in previous studies, constant monitoring of healthcare workers’ hand hygiene compliance and quality is crucial. In this study, we investigate the feasibility of using a thermal camera together with an RGB camera to detect hand coverage of alcohol-based formulation, thereby monitoring handrub quality. 
    \item[Materials and Methods:] We recruited 32 participants for this study. Participants were required to perform four types of handrubs to produce different hand coverage of alcohol-based formulation. After each task, participants' hands were photographed under a thermal camera and an RGB camera, while an ultraviolet (UV) test was used to provide the ground truth of hand coverage of alcohol-based formulation. Then, a U-Net was used to segment areas covered with alcohol-based formulations from thermal images, and the system performance was evaluated by comparing coverage differences between thermal images and UV images (from UV tests) regarding accuracy and Dice coefficient.
    \item[Results:] The system yields promising results in terms of accuracy (93.5\%) and Dice coefficient (87.1\%) when observations take place 10 seconds after performing handrub. In addition, we also examine the system performance change over a 60-second observation period, and the accuracy and Dice coefficient still remain at about 92.4\% and 85.7\% when observation happens at the 60-second time point.
    \item[Conclusions:] Given these encouraging results, thermal imaging shows its potential feasibility in providing accurate, constant, and systematic hand hygiene quality monitoring.
\end{description}

\end{abstract}







\makeatletter
\let\@authorsaddresses\@empty
\makeatother
\maketitle

\section{Introduction}
\label{sec:introduction}
Hand disinfection --- a major component of appropriate hand hygiene --- is the most effective way to prevent \ac{hais} and reduce their transmissions~\cite{Allegranzi2009RolePrevention, who2009guidelines, Pittet2001ImprovingApproach}. \ac{hais} are one of the most crucial patient-safety challenges in healthcare settings~\cite{Stefl2001To1999.}. They dramatically increase patients' length of stay, costs, morbidity, and mortality~\cite{Klevens2007Estimating2002, DeAngelis2010EstimatingCosts}. Moreover, \ac{hais} often lead to a heavy financial burden on healthcare systems. In the United States, the estimated annual costs range from \$28 billion to \$45 billion~\cite{Graves2009EconomicHospitals}. However, at least one-third of \ac{hais} could be prevented by following better hand hygiene practices~\cite{Stone2007EffectHygiene}.

In 2009, the \ac{who} issued the  ``\ac{who} guidelines on Hand Hygiene in Health Care'' providing a thorough review of evidence on hand hygiene in healthcare and specific recommendations to improve practices in healthcare settings~\cite{who2009guidelines}. In the guidelines, the \ac{who} summarises the five key moments when \ac{hcws} should perform hand hygiene~\cite{who2009guidelines}. The guidelines also recommend two standard hand hygiene procedures, including ``Handwash with Soap and Water'' for visibly soiled hands and ``Handrub with Alcohol-based Formulation'' for routine decontamination of hands for all clinical indications~\cite{who2009guidelines}.

However, research has found that hand hygiene quality in healthcare settings is generally unsatisfactory~\cite{Korhonen2015AdherenceFidelity, Arias2016AssessmentSteps, Szilagyi2013A6-steps}. Szilágyi \etal reported that only 72\% \ac{hcws} could adequately clean all hand surfaces immediately after hand hygiene training~\cite{Szilagyi2013A6-steps}. Since handrub with an alcohol-based formulation has been widely adopted into routine clinical practices~\cite{who2009guidelines}, precisely measuring handrub quality and providing \ac{hcws} with feedback regarding their performance are essential to promote good hand hygiene in healthcare environments.

Yet, measuring handrub quality in healthcare settings is challenging~\cite{Wang2021ElectronicTechnology}. Direct observation by trained auditors is considered the gold standard for monitoring both hand hygiene compliance and quality, but its rigor is limited by personnel time and expense, insufficient sample size, and the Hawthorne effect~\cite{Boyce2008HandUSA}. Recently, researchers measured hand hygiene quality by tracking \ac{hcws}' compliance with the \ac{who} 6-step hand hygiene procedures through environmental and wearable sensors~\cite{Korhonen2015AdherenceFidelity, Goroncy-Bermes2010ImpactHands, Llorca2011AAssessment, Hoey2010AutomatedProcess,Wang2020AccurateStudy, Kutafina2016WearableTraining, Li2018Wristwash:Device}; however, their techniques were proxy measures to detect \ac{hcws}' hand hygiene quality. Nevertheless, researchers attempted to quantify the hand coverage during handrub and handwash procedures~\cite{Wang2022ATechniques, wang2022hand}. \ac{uv} tests have been widely used for medical hand hygiene training to highlight insufficiently cleaned regions after \ac{hcws} handrubbing. However, since \ac{uv} tests require fluorescent dye, which often leaves residual which affects follow-up measurements,  and \ac{uv} lamps, it cannot be easily incorporated into \ac{hcws}' daily routines. 

Alcohol-based formulation typically contains 60\%-80\% ethanol, which vaporizes under room temperature and cools down hand surfaces. Given recent advances in thermal imaging, thermal cameras are now able to detect granular temperature changes, which strongly motivates our approach to use thermal imaging to measure the surface coverage of alcohol-based formulations.

In this paper, we compare the detected hand coverage of alcohol-based formulation between thermal images (collected by the thermal camera) and \ac{uv} images (RGB images collected from \ac{uv} tests as the ground truth) pixel by pixel. We use image registration to transform thermal images into the coordinate system of \ac{uv} images to calculate the estimated coverage difference in terms of accuracy and Dice coefficient (similarity between predicted and real clean areas). By comparison, we are able to explore the feasibility of using thermal cameras to measure hand hygiene quality by
\begin{enumerate}
    \item validating the accuracy of using thermal imaging to detect the coverage of alcohol-based formulation. The proposed thermal camera-based system yields the promising results of 93.5\% accuracy and 87.1\% Dice coefficient when observation happens at 10 seconds after performing handrub;
     \item measuring system performance changes over a 60-second observation since the heat signature degrades over time (although an observation can happen within a few hundred milliseconds). We find that the system performance decreases with progressing observation time, suggesting that the observation should be taken as soon as possible. Nevertheless, even after 60 seconds of observation time, the accuracy and Dice coefficient still remain at about 92.4\% and 85.7\%, respectively.
\end{enumerate}

\section{Method}
\subsection{Participants and Experimental Procedure}
We recruited 32 participants through our institution's mailing lists and snowball recruitment with an equal number of women and men. All participants were students or staff at our institution, and their ages ranged between 18 and 27 (M = 22.8, SD = 2.1 years). The majority of participants (27 out of 32, 84.4\%) had not received formal hand hygiene training in the last three years and were not familiar with the formal hand hygiene procedures. The entire experimental procedure lasted for about 130 minutes, including briefing and debriefing. This study was approved by the anonymous institution's Human Ethics Advisory Group. 

Before the experiment, participants completed a questionnaire about allergic reactions to alcohol-based formulation and \ac{uv} light. Upon arrival at our lab, we briefed participants on the study purpose and obtained their written consent agreeing to participate in our experiment. We subsequently provided training to our participants on how to perform the \ac{who} 6-step handrub procedures. To achieve this, we first explained the procedure steps. They then watched an instructional video provided by the  \ac{who}~\cite{who2015handrub} three times, and performed the handrub procedure along with the instructional video for training.

After the training session, participants proceeded to complete the experimental tasks. Before each task, participants rinsed their hands to remove residual fluorescent dye and dried their hands with tissues. Then, they rubbed their hands with hand warmers for one minute to rewarm their hands back to their body temperature~\cite{Oerlemans1999ReliabilityStudy}. We took both thermal and RGB images as baseline images before depositing alcohol-based formulation on participants' hands. After that, participants performed the experimental tasks and placed their hands on the pegboards for 60 seconds while they were observed by the thermal camera. Finally, the \ac{uv} lamp was turned on above the participants' hands, and a photograph was taken using the RGB camera. 

We designed 30 different tasks in line with four task types mentioned in Appendix Section~\ref{Sec:task-design} for both sides of participants' hands, as well as the task descriptions for each task and the visualized surface coverage examples. However, since participants are required to keep their hands on the pegboards during the 60-second observation period, this would extend the experiment duration up to 180 minutes in total, and potentially cause participant fatigue. Therefore, we reduce the experiment duration by only observing one side of each participant's hands (either palmar side or dorsal side) for the tasks of separated \ac{who} handrub steps. As a result, each participant completed 21 out of the 30 possible tasks. Specifically, tasks 5-13 (individual handrub steps, dorsal) and 20-28 (individual handrub steps, palmar) were completed by half the cohort. 

\subsection{Data Prepossessing}
\subsubsection{Image Registration}
\begin{figure}[H]
  \centering
  \includegraphics[width=.9\linewidth]{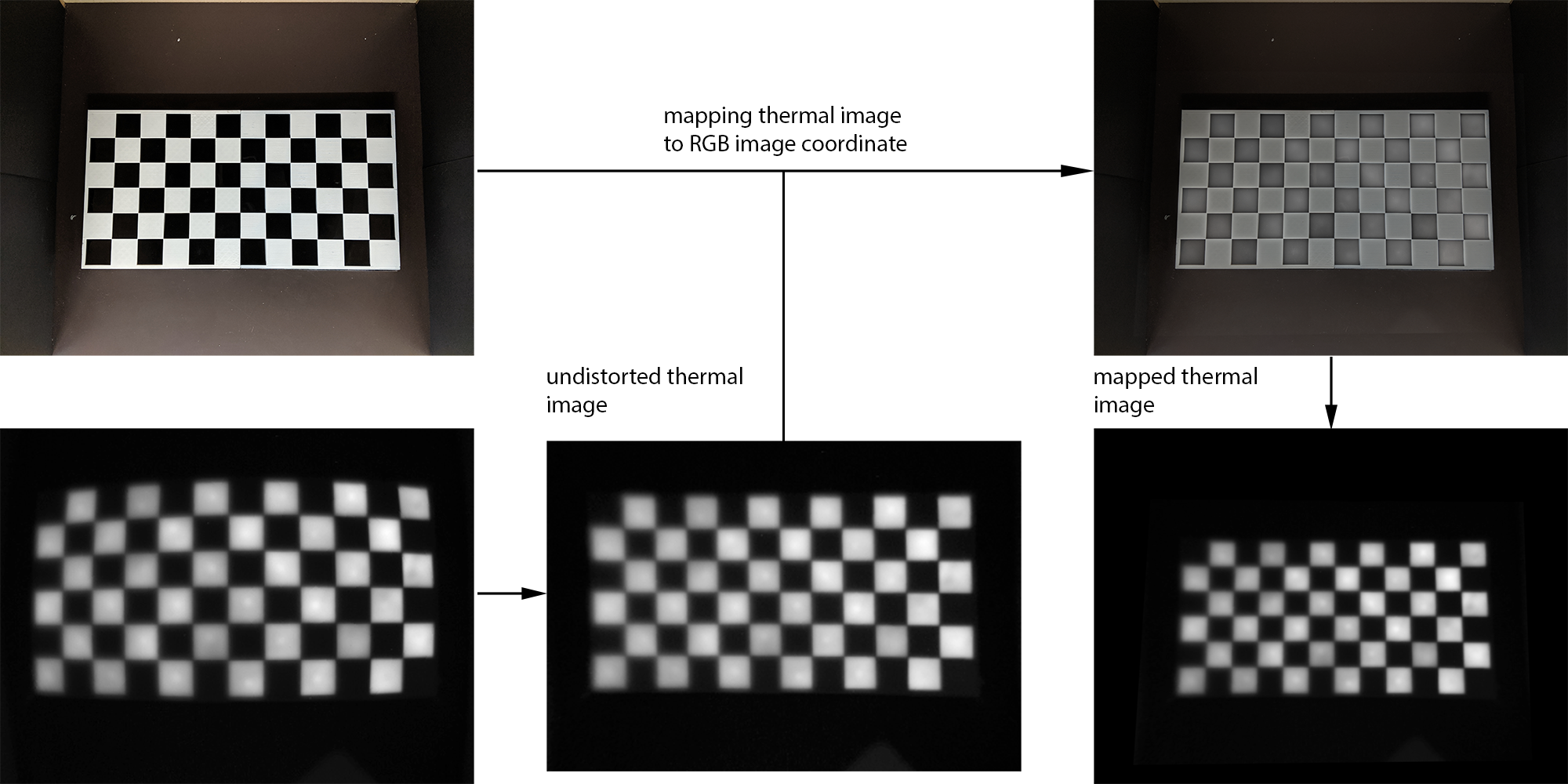}
  \caption{Procedure to register the thermal camera to the coordinate system of the RGB camera}
  \label{figure:ThermalChessboard}
\end{figure}

Our analysis requires that we transform the thermal and RGB images into one coordinate system. The first step is calibrating the thermal camera to correct its lens distortion. Camera calibration often uses chessboard or circle board patterns. However, in our case, simply printing a chessboard pattern is not adequate since it does not have thermal properties. Thus, we fabricated our $12 \times 6$ white chessboard using a 3D printer with 36 square pits, and we also fabricated 36 black cubes using a laser cutter. By placing the black cubes on the chessboard, the $12 \times 6$ chessboard pattern can be detected under the RGB camera (shown in Figure~\ref{figure:ThermalChessboard}). At the same time, these cubes can be heated using a microwave oven, and then placed on the chessboard, thus revealing a chessboard that can be seen with the thermal camera (shown in Figure~\ref{figure:ThermalChessboard}). After taking a set of thermal images of the chessboard, we further processed those images using the OpenCV library~\cite{Bradski2000TheLibrary}. We first extracted and recorded the positions of the internal corners of the chessboard from each image before computing the intrinsic parameters for the thermal camera to undistort the thermal images.

After correcting the lens distortion of the thermal camera, we then performed image registration between cameras. Since the resolution of the RGB images is much higher than the resolution of the thermal images, we chose to transfer the thermal images to the coordinate system of the RGB images to minimize the loss of details in the RGB images. First, we installed the cameras inside the wood frame (shown in Appendix Section~\ref{sec:device-setup}) in our experiment setup. Then, we heated the black cubes, installed the cubes on the chessboard, and simultaneously took an RGB image and a thermal image. We extracted the internal corners from both images and found their relative homography. In the end, we applied a perspective transformation to the thermal image into the coordinate system of the RGB image. To avoid errors introduced by the shift of cameras, we repeated the process of image registration between the cameras before each participant.

\subsubsection{Ground Truth from \ac{uv} Images}\label{sec:ground-truth}
After the camera calibration and registration, we marked the ground truths of the hand areas covered by alcohol-based formulation using \ac{uv} images. To recognize the hand areas from \ac{uv} images, we transferred the RGB images taken under white light into the YUV color system, where Y stands for luma component and U and V are the chrominance components. We then used the V channel (red-difference chroma component) and Otsu's method for automatic image thresholding~\cite{Otsu1979THRESHOLDHISTOGRAMS.}. After finding the largest contours inside the binary images, we could segment the hands from the background. Since this study mainly focuses on hand hygiene, we further segmented areas above the wrists from the detected hand as \ac{roi}, by manually labeling wrist points as ground truth.

Then, we extracted the alcohol-covered areas from the \ac{uv} images. Since the fluorescent dye used in the experiment glow in green light under the \ac{uv} lamp, we transferred these images to the \ac{hsv} color system and used the H channel to specifically detect areas within the green color range by using a threshold between 25 to 97. To minimize the impact of residual fluorescent dye from previous trials that are considered noise, we used the V channel to remove areas within the green color and with low brightness (under 60). 


\subsection{Segmentation of RGB and Thermal Images}
Here, we adopted a deep learning neural network -- U-Net, one of the most widely-used biomedical image segmentation algorithms, to segment (1) hand areas from RGB images and (2) covered areas from thermal images.

\begin{figure}[H]
    \centering
    \includegraphics[width=.9\linewidth]{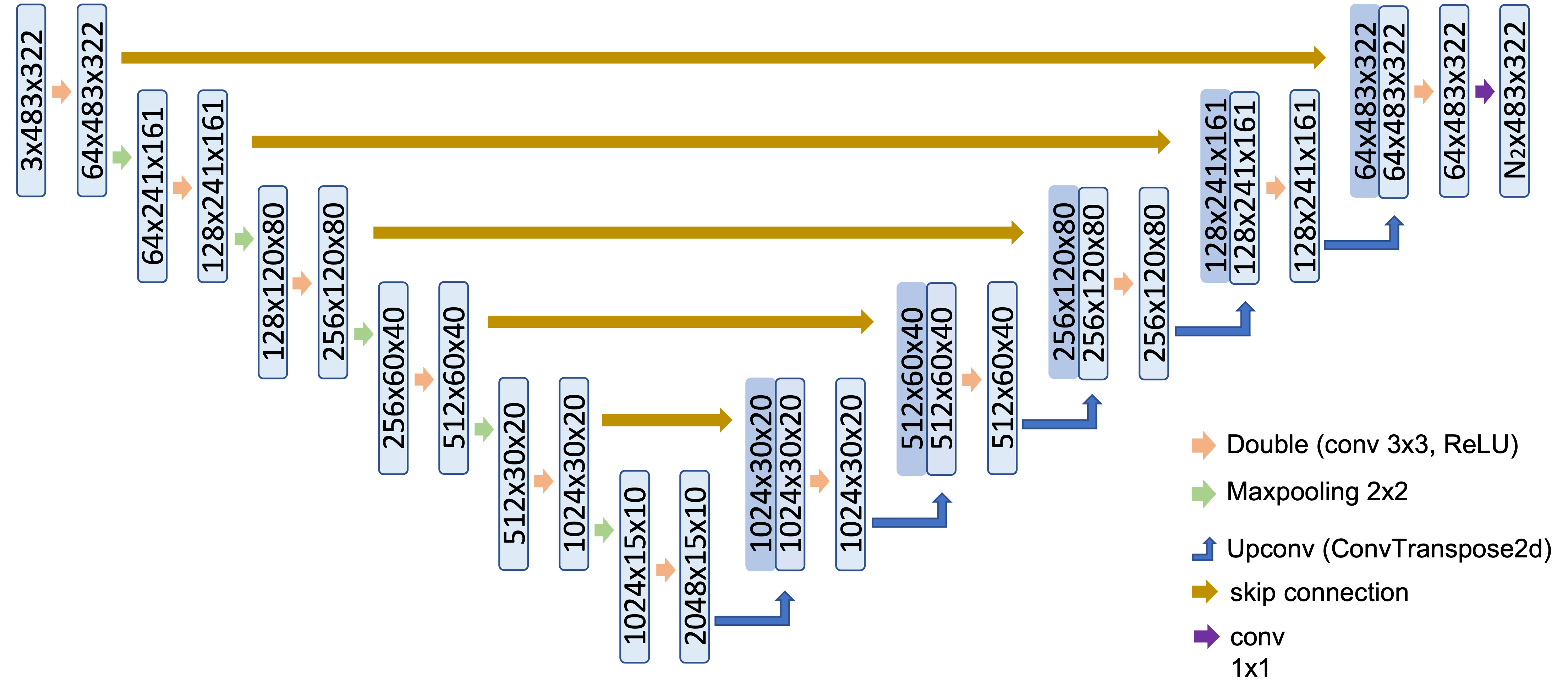}
    \caption{Model structure of the proposed models.}
    \label{fig:model-structure}
\end{figure}

\begin{figure}[H]
    \centering
    \includegraphics[width=.95\linewidth]{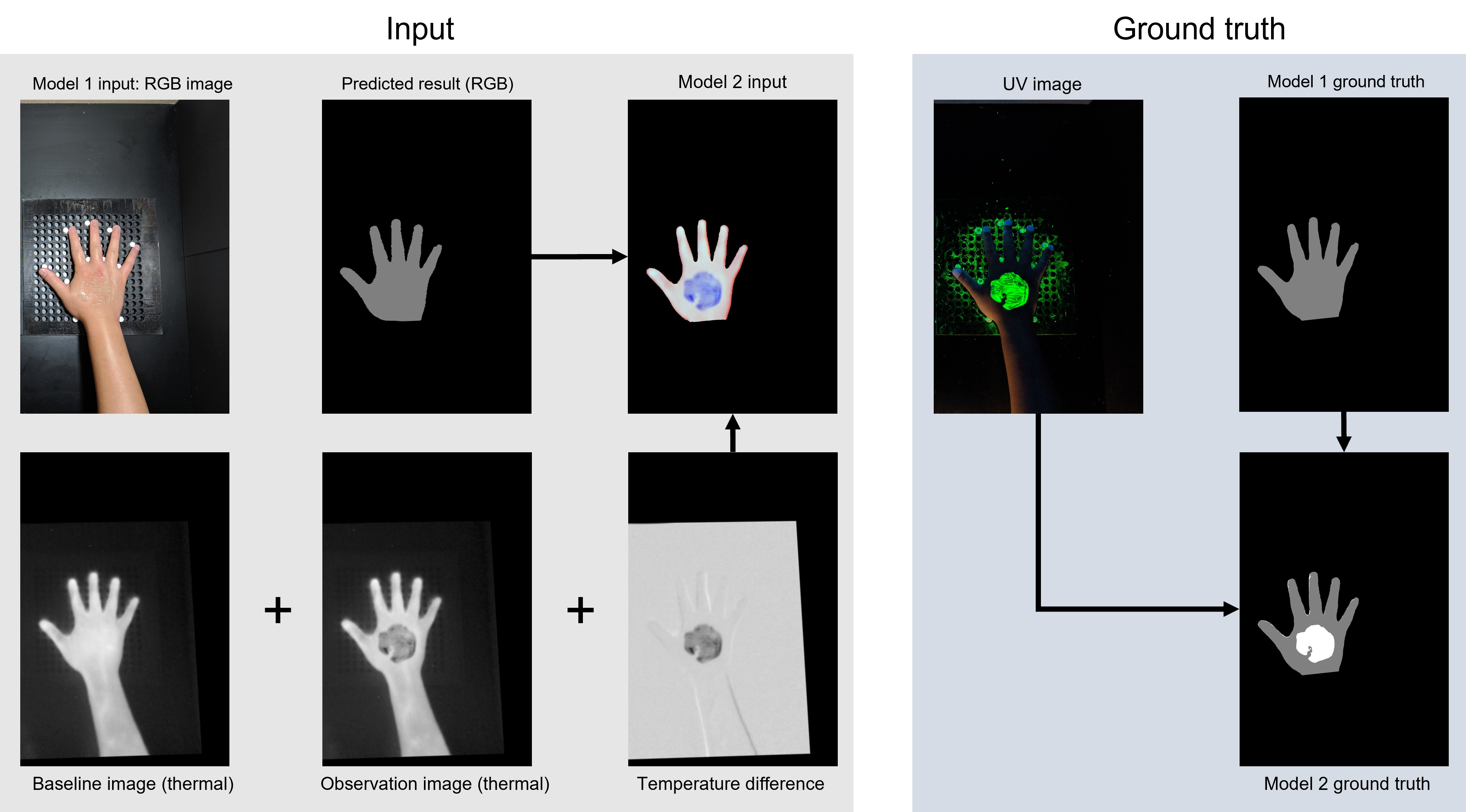}
    \caption{Required inputs of the proposed models.}
    \label{fig:model-input}
\end{figure}

We used the same network structure (shown in Figure~\ref{fig:model-structure}) for both segmentation tasks. The inputs to the model for task 1 (segment hand areas from RGB images) are shown in Figure~\ref{fig:model-input}. In the meantime, the inputs of task 2 (segment areas covered by alcohol-based handrub from thermal images) will be combined with the baseline image, observation image, and their differences. The background may be noisy when adopting thermal imaging in healthcare settings; therefore, we further employ the segmented hand areas generated by task 1 to remove the background information from the inputs of task 2. More details are shown in Figure~\ref{fig:model-input}. To increase the size of the dataset, we flipped images horizontally for data augmentation.

We trained both models with Combo loss by combining Cross-Entropy loss and Dice loss~\cite{taghanaki2019combo}. We implemented and trained the models in Pytorch with a single Nvidia GeForce RTX 3080 super (12GB RAM). Both models were trained for 30 epochs and a batch size of 8. We resized the input images to $3 \times 483 \times 322$ pixels (16\% of the original image). RMSprop optimization was used, with an initial learning rate of $10^{-5}$, weight decay of $10^{-8}$, and momentum of 0.9. We used the learning rate schedule on this basis: if the Dice coefficient on the validation set is not increased for two epochs, the learning rate will decay by a factor of 0.1. 

\subsection{Statistical Analyses}
The accuracy and Dice coefficient of both segmentation tasks were measured to assess the performance of the proposed systems, and 5-fold cross-validation was used to check their generalizability. For task 1, we compared the U-Net segmented hand areas with manually segmented hand areas. For task 2, since previous work has shown that fluorescent dye highlights the areas of the hand surface that are adequately disinfected with acceptable accuracy (95\% sensitivity and 98\% specificity), we compared the coverage difference between results detected by \ac{uv} images and thermal images.  More details are mentioned in Section~\ref{sec:ground-truth}~\cite{Lehotsky2017TowardsProcedure}.




For each image (including both left and right hands), we calculated accuracy across all classes (task 1: hand areas and background, and task 2: hand areas covered or uncovered with alcohol-based formulation). In the meantime, we calculated the Dice coefficient for \ac{roi}s (task 1: hand areas, and task 2: hand areas covered with alcohol-based formulation). However, since hands in several tasks will be either fully covered or not covered by alcohol-based formulation (\eg the task of Step 1 only; more details shown in Appendix Section~\ref{sec:coverage}), 0\% coverage will raise division by zero errors when calculating Dice coefficient. Therefore, we calculated $Dice \; coefficient = \dfrac{2 \times \ac{tp} + \epsilon}{2 \times \ac{tp}+\ac{fp}+\ac{fn} + \epsilon}$, where $\epsilon = 0.001$ to prevent division by zero errors.

Due to heteroscedasticity in the data, we adopted the Welch \ac{anova} and Welch-Satterthwaite degrees of freedom~\cite{Welch1967ThePeriodograms}. Statistical analyses were conducted using Python (version 3.6.8) and statsmodels (version 0.9.0).

\section{Results}
\subsection{Segmentation Performance}
We discarded the data of 3 participants (Participants 1, 13, and 22) because they did not precisely follow the study protocol (\eg failed to wash out residual \ac{uv} dye, or did not keep their hands still during the observation period). Considering the accuracy of the system, it is important to note that the timing of measurements matters. As the alcohol evaporates from participants' hands, it causes a temporary temperature drop. This means that if the thermal observation happens a long time after the alcohol is applied, there may be no observable effect. Given these constraints, we report the accuracy and Dice coefficient at the 10-second observation time: the thermal imaging observation happens 10 seconds after participants place their hands on the pegboards, due to its highest accuracy. In the subsequent section, we report the effect of increasing this time window to up to 60 seconds. 

\begin{figure}[H]
    \centering
    \includegraphics[width=\linewidth]{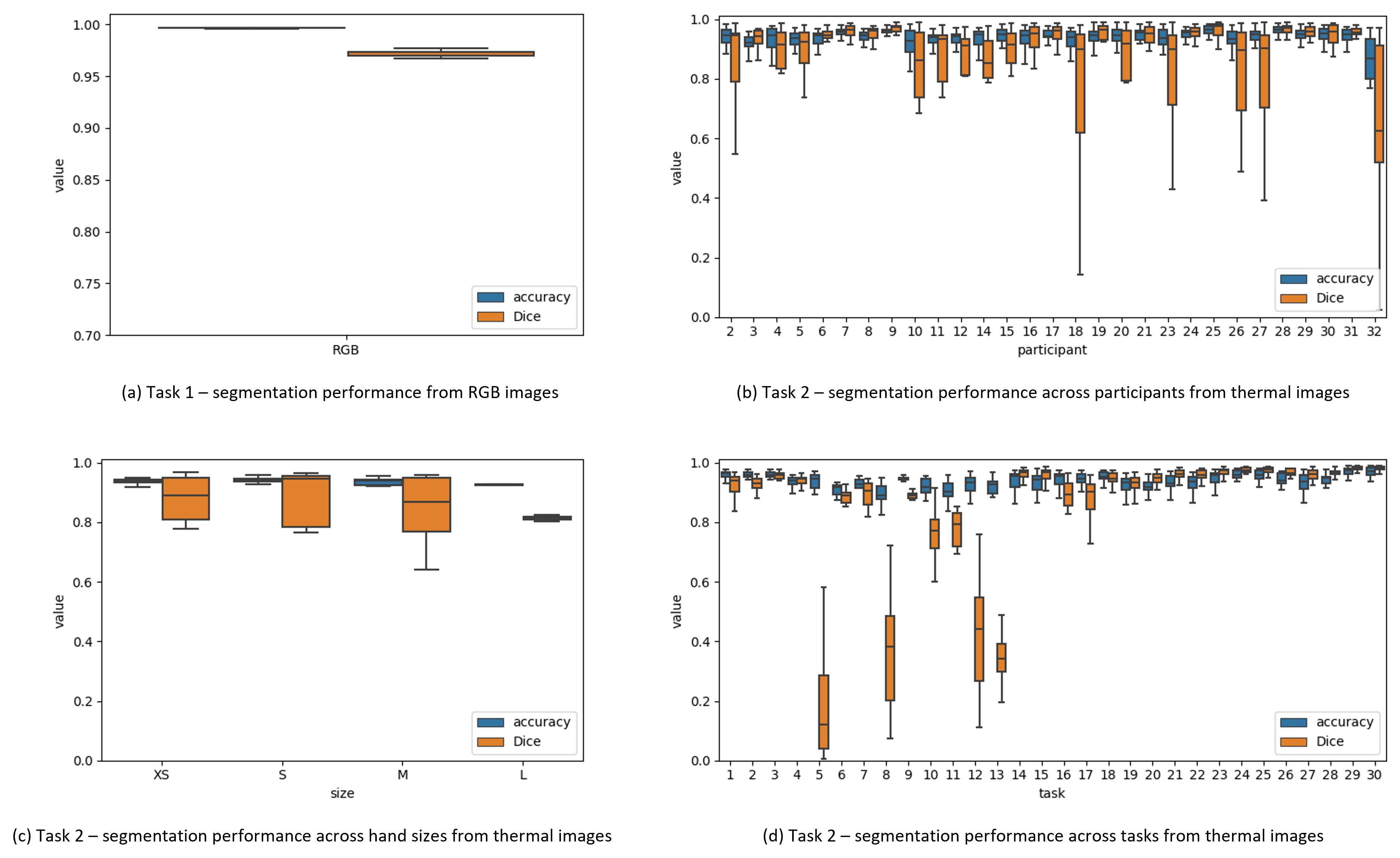}
    \caption{Classification accuracy and Dice coefficient of participants, hand sizes, and tasks.}
    \label{fig:participant-experiment-accuracy}
\end{figure}

For the model of task 1, the mean accuracy and Dice coefficients throughout 5-fold cross-validation are 99.6\% ($SD=0.0003$) and 97.2\% ($SD = 0.003$), respectively. More details are shown in Figure~\ref{fig:participant-experiment-accuracy}a. Meanwhile, to validate the accuracy of using thermal imaging to detect the coverage of alcohol-based formulation, we summarize the performance (in terms of accuracy and Dice coefficient) of thermal imaging across participants, hand sizes, and tasks (coverage ranges from 0\% to 100\%) to ensure the reliability and validity of the results. 

The system recognizes the hand areas covered by alcohol-based formulation with a mean accuracy of 93.5\% ($SD = 0.046$) and a mean Dice coefficient of 87.1\% ($SD = 0.195$) for all participants across all the experiments. We also group the accuracy for each participant for different tasks and present the results in Figure~\ref{fig:participant-experiment-accuracy}b. Of these, the highest mean accuracy is 96.0\% for Participant 25, and the lowest mean accuracy is 84.9\% for Participant 32. In the meantime, Participant 7 has the highest mean Dice coefficient (96.0\%), and Participant 32 has the lowest mean Dice coefficient (64.0\%). 

Furthermore, we measure the effect of hand size on accuracy. Because we used a fixed amount of alcohol per participant task (3 ml as recommended by ~\cite{Boyce2002GuidelineForce}), there will be less alcohol per unit area for participants with larger hands. Thus, we group participants in terms of hand length: $160 \le XS < 171 mm$, $171 \le S < 182 mm$, $182 \le M < 192 mm$, and $192 \le L < 204 mm$, and each participant's mean accuracy is considered as one data point (shown in Figure~\ref{fig:participant-experiment-accuracy}c). A one-way \ac{anova} does not show a significant difference in system accuracy for different hand size groups ($F_{3, 25} = 1.302, P = 0.295$) as well as in Dice coefficient ($F_{3, 25} = 0.662, P = 0.583$).

We also measure the mean accuracy for each task by summarizing results across all participants (shown in Figure~\ref{fig:participant-experiment-accuracy}d). Of these, Task 30 has the highest mean accuracy of 96.3\%, and Task 8 has the lowest mean accuracy of 88.7\%. Furthermore, Task 30 has the highest mean Dice coefficient of 97.7\%, and Task 5 has the lowest mean Dice coefficient of 17.6\%. 

As seen in the aforementioned results, Dice coefficients show several dramatic drops for some participants and tasks even though their accuracy is still above 90\%. This phenomenon may associate with the small sizes of the areas that alcohol-based formulations cover. Therefore, we applied Spearman's Correlation test to examine the correlation between Dice coefficients and sizes of the areas covered by alcohol-based formulations. For each participant, the mean percentage of areas covered with alcohol-based formulations across all tasks varies from 34.0\% to 73.6\%, and there exists a strong positive correlation with Dice coefficients (0.840, $P < 0.001$). For each task, the mean percentage of areas covered with alcohol-based formulations across all participants varies from 6.1\% to 94.2\%, and there is a strong positive correlation with Dice coefficients (0.967, $P < 0.001$) as well. These findings suggest that because of the limited resolution of thermal cameras and the imperfect alignment between baseline and observation images, thermal imaging may not be able to identify areas with small sizes and near edges.

\subsection{Effects of Varying Observation}
\begin{figure}[H]
    \centering
    \includegraphics[width=.85\linewidth]{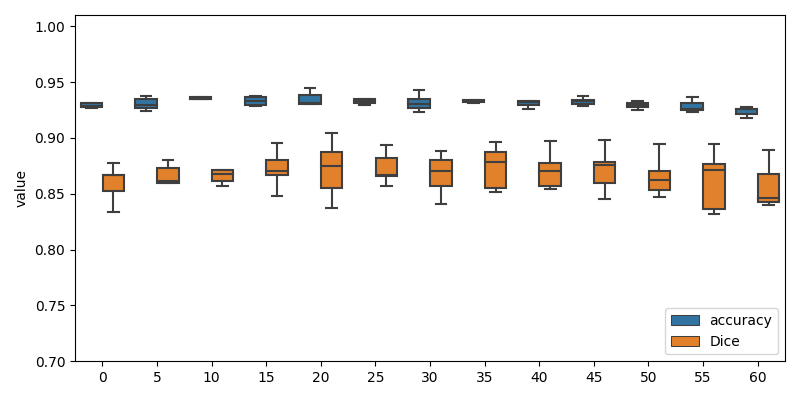}
    \caption{System performance of accuracy and Dice coefficient across 60 seconds observation.}
    \label{fig:duration-metrics}
\end{figure}

In the experiment, participants were required to place their hands on the observation pegboards for 60 seconds after completing all 21 tasks. Throughout this 60-second observation period, we measured the effect of the gradual hand rewarming on the system performance. In this section, accuracy and Dice coefficient are calculated for each participant and then grouped by the observation time across all 30 participants. Due to the computational requirements of the analysis, we analyze the thermal imaging results in every 5 seconds from 0 delay (images immediately captured after participants place their hands on the observation pegboards) up to a 60-second delay, as shown in Figure~\ref{fig:duration-metrics}.

The maximum mean accuracy of 93.6\% is displayed at 10 seconds, where all mean accuracy values are above 92\% across the 60-second observation period. Meanwhile, all mean Dice coefficients are above 85\% between 0 seconds and 60 seconds, with the highest mean Dice coefficient of 87.4\% occurring at 35 seconds. More details are shown in Figure~\ref{fig:duration-metrics}.

Across the 60-second observation period, accuracy and Dice coefficient gradually decrease slightly over time. Therefore, we evaluate the correlation between the observation time and system performance through Spearman's correlation test. The correlation values between the observation time and accuracy and Dice coefficients are -0.275 ($P = 0.026$) and -0.034 ($P = 0.785$), respectively. Although it is recommended to collect thermal images as soon as possible because of the weakly negative correlation between observation time and system performance, thermal imaging continues to operate effectively during the 60-second observation period.

\section{Discussion}
\subsection{Performance of Thermal Imaging}
In our study, thermal imaging presents promising results in observing hand hygiene quality in a laboratory setting. By comparing the detected areas between thermal images and \ac{uv} images, the thermal camera-based system achieves the highest comprehensive performance at the 10-second observation time with an accuracy of 93.5\% and a Dice coefficient of 87.1\%, which means the system could recognize hand coverage alike the gold standard of using fluorescent dye and \ac{uv} light~\cite{Reilly2016AKingdom, Lehotsky2017TowardsProcedure}. 

However, a major disadvantage of fluorescent dye is that it is impractical due to the residue it leaves, and it is typically only used in educational settings. Compared to the \ac{uv} test, thermal imaging recognizes hand coverage by measuring granular temperature changes caused by alcohol-based formulation without adding extra components. Nonetheless, thermal imaging may be unable to detect areas covered by alcohol-based formulations that are small and close to edges, due to the low resolution of thermal cameras and the imperfect alignment between baseline photos and observation images. This shortcoming necessitates the development of more advanced image segmentation and registration algorithms in order to segment hands from the background and areas covered with alcohol-based formulation and align hands between baseline and observation images.

Also, as the observation time increases and the handrub slowly dissipates, the system performance gradually declines. Even though the observation time of 60 seconds has the lowest accuracy and Dice coefficient, the values for both accuracy and Dice coefficient remain at 92.4\% and 85.7\%, respectively. This means that thermal imaging is reasonably accurate in recognizing the majority part of the hand surfaces covered and uncovered by the alcohol-based formulation.

Our study demonstrates how thermal imaging could potentially benefit \ac{hcws} in their hand hygiene activities. This is particularly important during the COVID-19 pandemic and post-pandemic times, as hand hygiene has been highlighted as one of the most efficient ways to prevent the transmission of viruses and \ac{hais}. However, due to the low compliance and quality of hand hygiene reported in previous studies~\cite{Pittet2004HandPerceptions, Erasmus2010SystematicCare, Korhonen2015AdherenceFidelity, Arias2016AssessmentSteps, Szilagyi2013A6-steps}, constant monitoring of \ac{hcws}' hand hygiene compliance and quality is necessary. By adopting a thermal camera-based system in healthcare settings, \ac{hcws} and hospital managers could continuously monitor hand hygiene quality without having to interrupt their daily routines.


\subsection{Adopting Thermal Imaging in Healthcare Settings}
\begin{figure}[H]
    \centering
    \includegraphics[width=\linewidth]{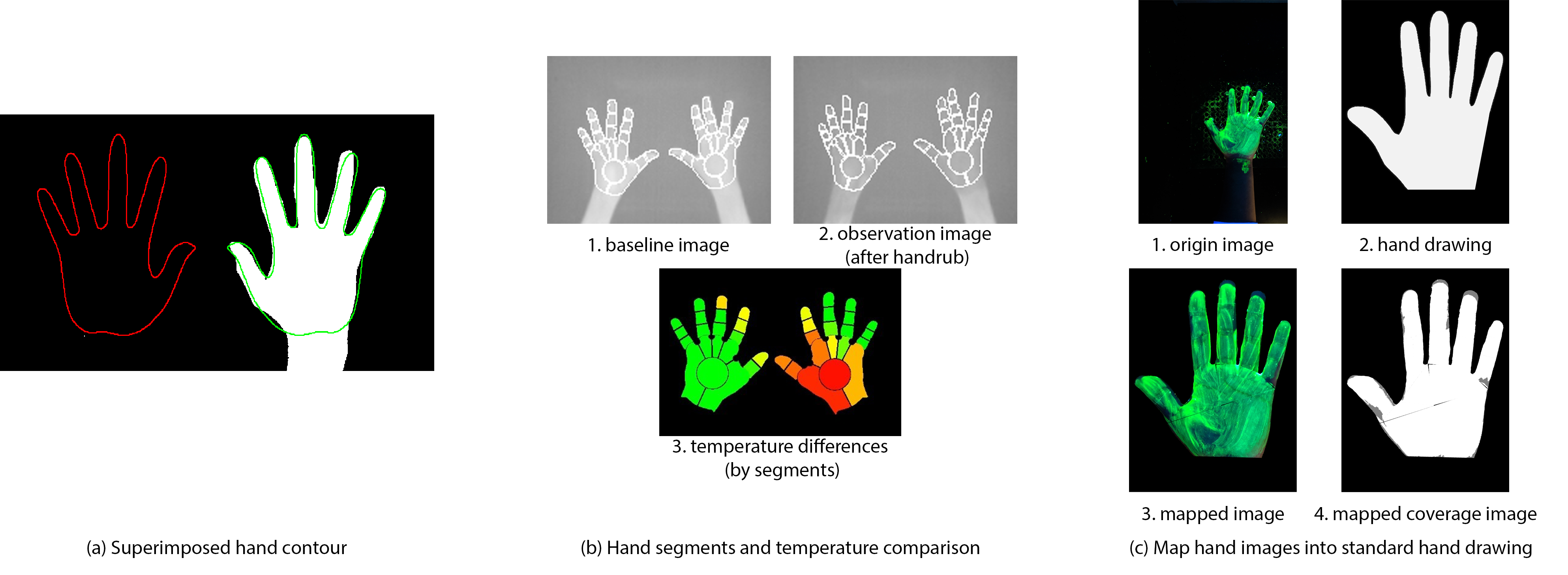}
    \caption{Alternative approaches for deploying thermal imaging in healthcare settings. Approach (a) is based on \cite{Smieschek2019AidedVision}, Approach (b) is based on \cite{Smieschek2016EvaluatingImaging}, and Approach (c) is based on \cite{Wang2022ATechniques}.}
    \label{fig:solution-adopt}
\end{figure}

Given that it is impractical to preserve users' hand positions between two observations, we suggest three alternative adoption approaches for thermal imaging in healthcare settings to register between baseline and observation images (shown in Figure~\ref{fig:solution-adopt}). 

The first approach is to provide a \ac{gui}, which shows a wire-frame RGB image with superimposed hand contours. The \ac{gui} would instruct \ac{hcws} to place their hands at the same position when taking baseline images and observation images after handrubbing (details shown in Figure~\ref{fig:solution-adopt}~\cite{Smieschek2019AidedVision}). However, it may run into the issue of hand misalignment, which might result in classification mistakes and poorer system performance. 

The second approach is to section hands into segments and then compare the temperature differences of the same segments between the baseline images and observation images (details shown in Figure~\ref{fig:solution-adopt}~\cite{Smieschek2016EvaluatingImaging}). While \ac{hcws} are aware that a certain segment is exposed to alcohol-based formulation, they are unable to identify which parts within the segment are absent due to the lack of precise coverage information that comes with this solution. 

The third approach is to avoid calculating the temperature differences for segments, but rather map hand segments to a standard hand drawing (details shown in Figure~\ref{fig:solution-adopt}~\cite{Wang2022ATechniques}). The system first splits recognized hand regions and hand drawings into 18 segments based on the landmarks generated by MediaPipe and finger-web points~\cite{Lugaresi2019MediaPipe:Pipelines}. Then, the segments of the hand regions are matched and mapped to the corresponding segments of the hand drawings. After that, the system can calculate temperature differences without losing coverage information. As a result, \ac{hcws} can recognize missed areas, and the visual intervention could help them improve their hand hygiene. This approach may offer extensive information inside hand segments and is more resilient to hand misalignment than the other two aforementioned approaches. 

\section{Conclusion}
In this paper, we showed the feasibility of using thermal imaging to detect hand coverage with alcohol-based formulation, thereby monitoring hand hygiene quality. In an evaluation with 32 participants, the system achieved promising results in terms of accuracy and Dice coefficient while being comparable to the gold standard for \ac{uv} concentrate. Our study shows the potential flexibility of employing thermal imaging to monitor hand hygiene quality, which can be a step toward a continuous automated hand hygiene monitoring system that allows real-time monitoring without interrupting the \ac{hcws}' daily routines.

\subsection*{Conflict of interest}
None declared.

\subsection*{Acknowledgments}
This work is partially funded by NHMRC grants 1170937 and 2004316, European Union’s Horizon 2020 research and innovation program (Grant agreement 957296), and PhD Write Up Award from Faculty of Engineering and Information Technology, the University of Melbourne. 

\bibliographystyle{ACM-Reference-Format}
\bibliography{main}

\newpage
\section*{Figures}

\clearpage

\newpage
\appendix

\setcounter{table}{0}
\captionsetup[table]{name=Appendix Table}
\setcounter{figure}{0}
\captionsetup[figure]{name=Appendix Fig.}
\setcounter{page}{1}

\section{Hardware Setup}\label{sec:device-setup}
We collected RGB images using an RGB camera from a smartphone (Pixel 2XL, Google LLC) with a resolution of $4032 \times 3024$ pixels. Thermal images were collected by a thermal camera (Optris Xi 400, Optris GmbH) with a resolution of $382 \times 288$ pixels, a field of view of $18^{\circ} \times 14^{\circ}\,(f = 20\,mm)$, and a thermal sensitivity (NETD) of 80 mK. We mounted the RGB and thermal cameras centrally on a wood frame and placed them adjacent to each other (details shown in Appendix Figure~\ref{figure:DeviceSetup}).

\begin{figure}[h]
  \centering
  \includegraphics[width=\linewidth]{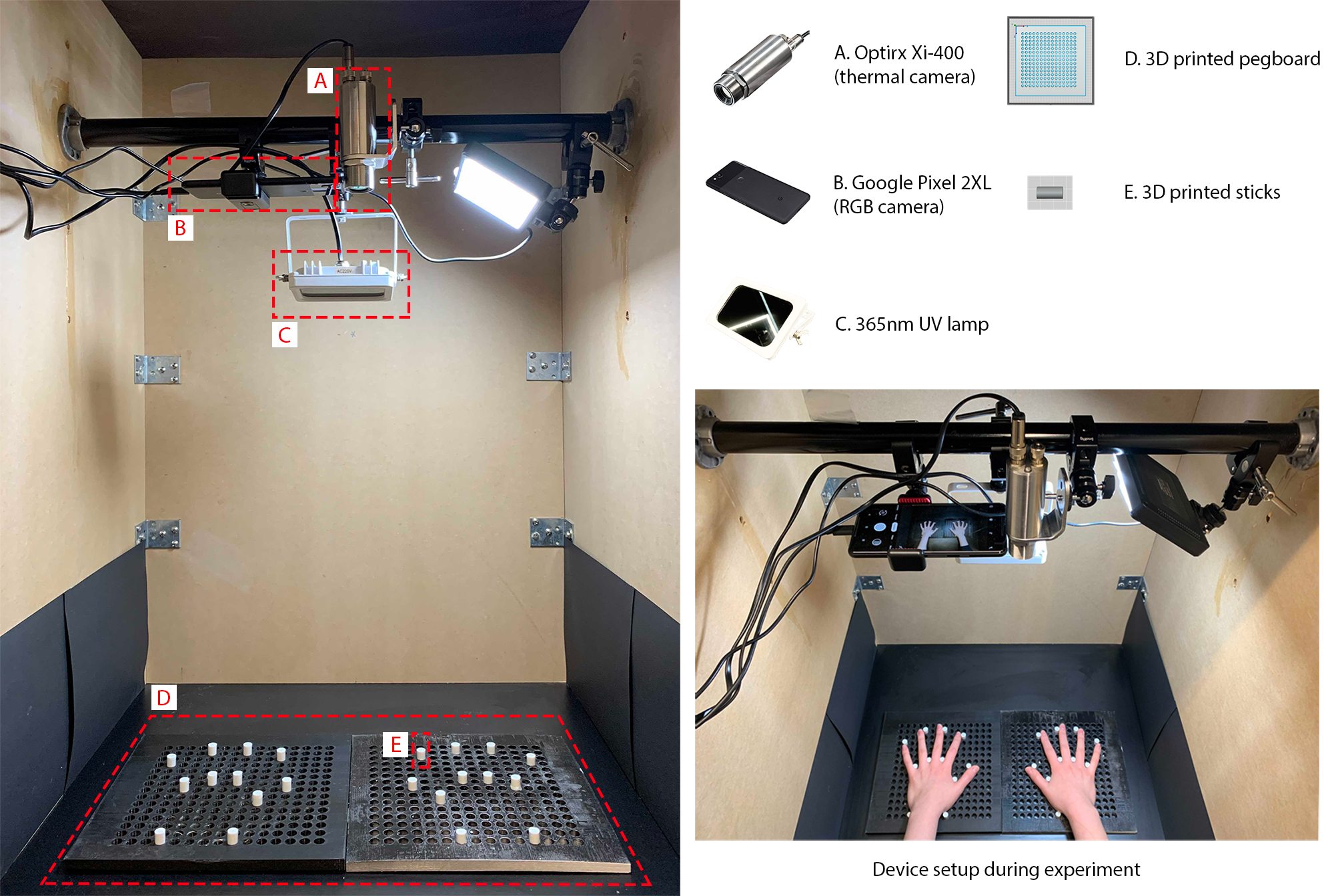}
  \caption{Experimental setup.}
  \label{figure:DeviceSetup}
\end{figure}

Our objective is to measure the accuracy of using a thermal camera to estimate the area size where alcohol-based formulation is spread on participants' hands. We consider the \ac{uv} images (RGB images taken under the \ac{uv} lamp) to represent ground truth. To validate the accuracy of using thermal imaging, we calculate the coverage differences between \ac{uv} images versus thermal images. This means that both cameras have to take a photo of the same sample (\ie participants' hands). To make this possible, we mixed a coloured antimicrobial hand gel (Microshield Angel Blue, Schülke \& Mayr GmbH) with a fluorescent handrub formulation (Glitterbug Gel, OnSolution Pty Ltd) using a ratio of 3:2. This means that the alcohol-based formulation maintains its thermal attributes, and simultaneously it has fluorescent properties making it visible under \ac{uv} light. Also, to ensure participants' safety, we use a 365 nm \ac{uv}-A lamp with an effective irradiance of 30 $mW / cm^2$ to detect the coverage of the fluorescent handrub formulation. The distance between the \ac{uv}-A lamp and the pegboards is 55 cm. We also control each participant's total observation duration to be below the daily \ac{uv} exposure threshold limit of 3.0 $mJ / cm^2$~\cite{ACGIH2016DocumentationEd.}. 

Moreover, we expect participants' hands to begin warming up after the alcohol has evaporated from their hands, which may degrade the usefulness of thermal imaging. Thus, we continuously record the performance changes to the thermal imaging over a period of 60 seconds. However, during this 60-second observation, participants may move their hands, resulting in misalignment when calculating coverage differences between two thermal images. To minimise participants' hand movements during the observation, we created a pair of pegboards and sticks using a 3D printer that help participants keep their hands in one spot (shown in Appendix Figure~\ref{figure:DeviceSetup}). We note that in an actual deployment, pegboards and sticks are unnecessary since they are designed to facilitate the 60-second observation to measure system performance changes, but an assessment can happen within a few hundred milliseconds.

\section{Task Design}~\label{Sec:task-design}
We designed four types of tasks that use fluorescent handrub formulation to cover participants' hands with different area sizes and positions on the hand to systematically evaluate thermal imaging performance. The tasks include concise patterns that can be easily recognized and processed by image processing software. Also, we use the \ac{who} recommended handrub procedure as two of the tasks~\cite{who2009guidelines}. The \ac{who} handrub procedure contains six steps and each individual step is also summarized in Appendix Table~\ref{table:HandHygieneSteps}), with some steps being repeated for each hand. 

\begin{table}[h]

\caption{Steps of the World Health Organization 6-step handrub procedure.}
\label{table:HandHygieneSteps}

\begin{threeparttable}
    
    
    \begin{tabular}{|l|l|}
    \hline
    Step\# & Description of Individual Step                                                             \\ \hline
    1      & Rub hands palm to palm                                                                     \\ \hline
    2 (R\tnote{a}\enspace)  & Right palm over left dorsum with interlaced fingers                                        \\ \hline
    2 (L\tnote{b}\enspace)  & Left palm over right dorsum with interlaced fingers                                        \\ \hline
    3      & Palm to palm with fingers interlaced                                                       \\ \hline
    4      & Backs of fingers to opposing palms with fingers interlocked                                \\ \hline
    5 (R)  & Rotational rubbing of left thumb clasped in right palm                                     \\ \hline
    5 (L)  & Rotational rubbing of right thumb clasped in left palm                                     \\ \hline
    6 (R)  & Rotational rubbing, backwards and forwards with clasped fingers of right hand in left palm \\ \hline
    6 (L)  & Rotational rubbing, backwards and forwards with clasped fingers of left hand in right palm \\ \hline
    \end{tabular}

    \begin{tablenotes}
        \item[a] R: right.
        \item[b] L: left.
    \end{tablenotes}
  
\end{threeparttable}

\end{table}

Thus, participants have to complete the following types of tasks in the experiment (task description mentioned in Appendix Section~\ref{sec:task-description} and visualized examples shown in Appendix Section~\ref{sec:coverage}):
\begin{enumerate*}
     \item Shapes -- We imprint the alcohol-based formulation on participants' hands using four stamps with different shapes, including circle, hexagon, square, and star.
      \item Equal Split -- We use a brush to apply a layer of the alcohol-based formulation onto participants' hands with 0\%, 50\%, and 100\% coverage.
     \item Individual \ac{who} Handrub Steps -- Participants are asked to perform either Step 1 alone, or Step 1 followed by each individual \ac{who} handrub steps described in Appendix Table~\ref{table:HandHygieneSteps} separately, using 3 ml alcohol-based formulation~\cite{who2009guidelines}. Step 1 is needed before performing other individual steps, because other steps cannot spread alcohol-based formulation without Step 1.
     \item Entire \ac{who} Handrub Procedure -- Participants are asked to perform the entire \ac{who} handrub procedure using 3 ml alcohol-based formulation~\cite{who2009guidelines}.
\end{enumerate*}

\subsection{Task description for the experiment}
\label{sec:task-description}
\begin{table}[H]

\caption{Task description for the experiment.}
\label{table:experiment_tasks}

\begin{threeparttable}

\begin{tabular}{|c|c|c|c|}
\hline
\textbf{Task} & \textbf{Hand Side} & \textbf{Task Type}         & \textbf{Description}                        \\ \hline
1             & Dorsal             & Stamp                      & Circle  (L\tnote{a}\enspace) + Hexagon    (R\tnote{b}\enspace)                  \\ \hline
2             & Dorsal             & Stamp                      & Rectangle (L) + Star   (R)                  \\ \hline
3             & Dorsal             & Brush                      & 0\% (L) + 100\% (R)                         \\ \hline
4             & Dorsal             & Brush                      & 50\%   (L, Top Half) + 50\% (R, Lower Half) \\ \hline
5             & Dorsal             & Individual Handrub   Steps & 1                                           \\ \hline
6             & Dorsal             & Individual Handrub   Steps & 1 + 2 (R)                                   \\ \hline
7             & Dorsal             & Individual Handrub   Steps & 1 + 2 (L)                                   \\ \hline
8             & Dorsal             & Individual Handrub   Steps & 1 + 3                                       \\ \hline
9             & Dorsal             & Individual Handrub   Steps & 1 + 4                                       \\ \hline
10            & Dorsal             & Individual Handrub   Steps & 1 + 5 (R)                                   \\ \hline
11            & Dorsal             & Individual Handrub   Steps & 1 + 5 (L)                                   \\ \hline
12            & Dorsal             & Individual Handrub   Steps & 1 + 6 (R)                                   \\ \hline
13            & Dorsal             & Individual Handrub   Steps & 1 + 6 (L)                                   \\ \hline
14            & Dorsal             & Entire Handrub   Procedure & -                                           \\ \hline
15            & Dorsal             & Entire Handrub   Procedure & -                                           \\ \hline
16            & Palmar             & Stamp                      & Hexagon (L) + Circle   (R)                  \\ \hline
17            & Palmar             & Stamp                      & Star (L) + Rectangle   (R)                  \\ \hline
18            & Palmar             & Brush                      & 100\% (L) + 0\% (R)                         \\ \hline
19            & Palmar             & Brush                      & 50\%   (L, Lower Half) + 50\% (R, Top Half) \\ \hline
20            & Palmar             & Individual Handrub   Steps & 1                                           \\ \hline
21            & Palmar             & Individual Handrub   Steps & 1 + 2 (R)                                   \\ \hline
22            & Palmar             & Individual Handrub   Steps & 1 + 2 (L)                                   \\ \hline
23            & Palmar             & Individual Handrub   Steps & 1 + 3                                       \\ \hline
24            & Palmar             & Individual Handrub   Steps & 1 + 4                                       \\ \hline
25            & Palmar             & Individual Handrub   Steps & 1 + 5 (R)                                   \\ \hline
26            & Palmar             & Individual Handrub   Steps & 1 + 5 (L)                                   \\ \hline
27            & Palmar             & Individual Handrub   Steps & 1 + 6 (R)                                   \\ \hline
28            & Palmar             & Individual Handrub   Steps & 1 + 6 (L)                                   \\ \hline
29            & Palmar             & Entire Handrub   Procedure & -                                           \\ \hline
30            & Palmar             & Entire Handrub   Procedure & -                                           \\ \hline
\end{tabular}

\begin{tablenotes}
    \item[a] L: left.
    \item[b] R: right.
\end{tablenotes}
  
\end{threeparttable}

\end{table}

\newpage
\subsection{Surface areas covered by alcohol handrub}
\label{sec:coverage}
\begin{figure}[h]
    \centering
    \includegraphics[width=\linewidth]{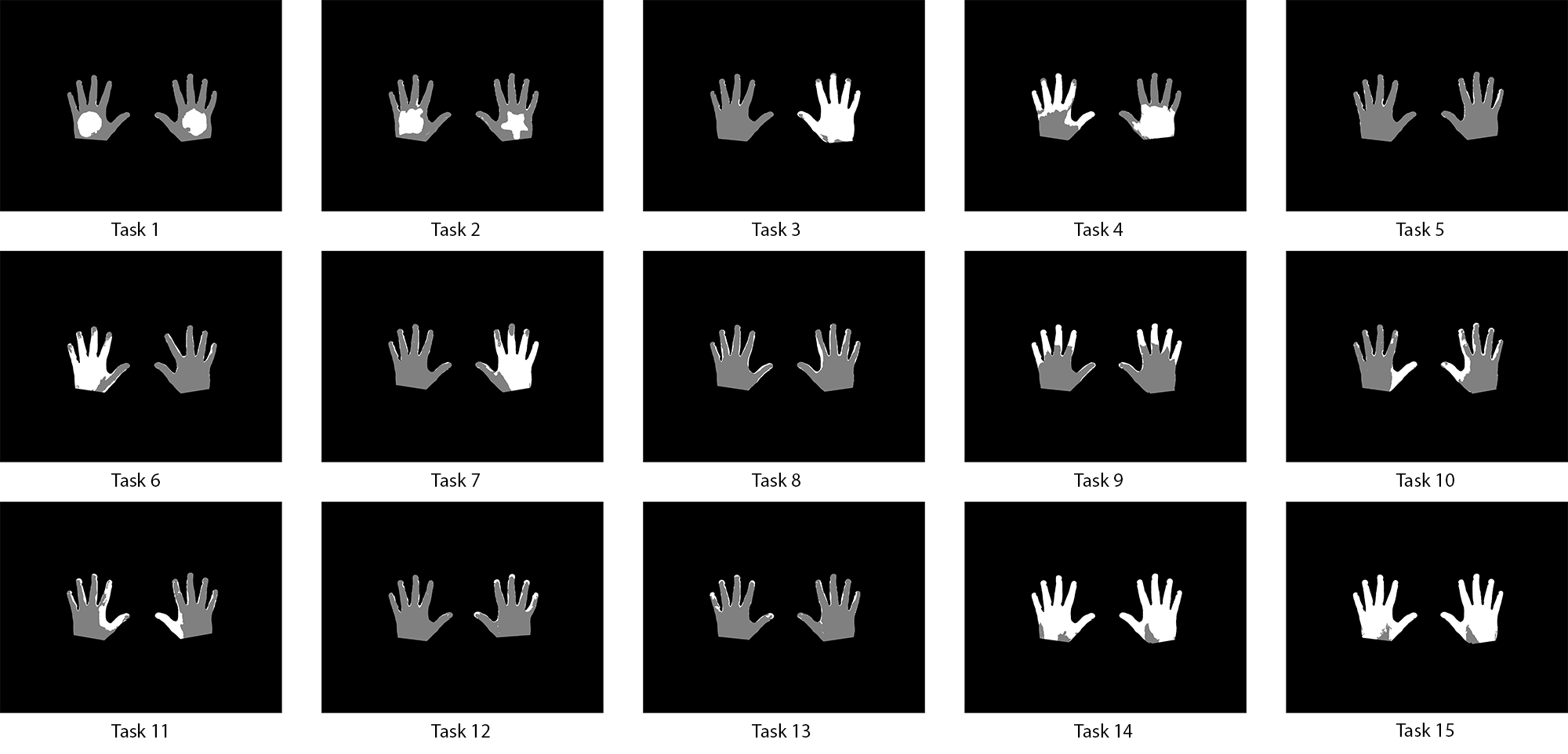}
    \caption{Surface areas covered by alcohol handrub for Task 1-15 (dorsal side).}
    \label{fig:dorasal_coverage}
\end{figure}

\begin{figure}[h]
    \centering
    \includegraphics[width=\linewidth]{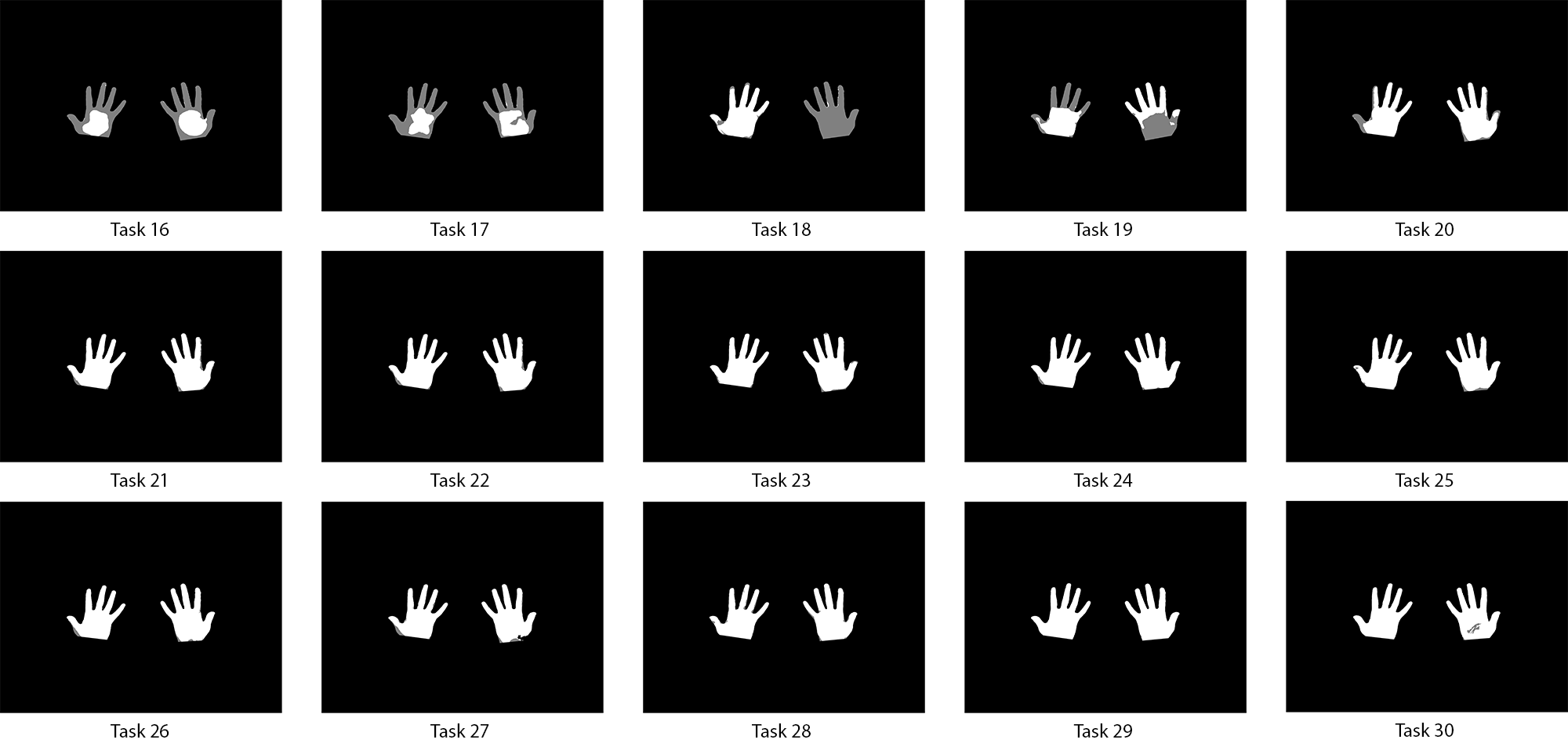}
    \caption{Surface areas covered by alcohol handrub for Task 16-30 (palmar side).}
    \label{fig:palmar_coverage}
\end{figure}


\end{document}